%% file: Held.tex
\documentclass[runningheads,a4paper]{llncs}

\usepackage{cite}

\usepackage{amssymb}
\usepackage{graphicx}

\usepackage{url}

\graphicspath{{pics/}}
\usepackage[cmex10]{amsmath}
\usepackage{icomma}

\usepackage{algorithmic}
\usepackage{url}

\usepackage[bookmarks=false]{hyperref}

\usepackage{booktabs}
\usepackage[utf8]{inputenc}
\usepackage[T1]{fontenc}

\newcommand{\argmax}{\operatornamewithlimits{argmax}}

\usepackage{units}

\begin{document}

\mainmatter  

\title{Online Community Detection\\by Using Nearest Hubs}

\titlerunning{Online Community Detection by Using Nearest Hubs}

%
%
\author{Pascal Held\and Rudolf Kruse}
\authorrunning{P. Held, R. Kruse}

\institute{Otto von Guericke University of Magdeburg\\
Department of Knowledge Processing and Language Engineering\\
Universit\"atsplatz 2, 39106 Magdeburg, GERMANY\\
\url{pascal.held@ovgu.de}\\
\url{rudolf.kruse@ovgu.de}\\
\url{http://fuzzy.cs.uni-magdeburg.de}}

%
%

\toctitle{Lecture Notes in Computer Science}
\tocauthor{Authors' Instructions}
\maketitle

\begin{abstract}
\input{chapters/0_abstract}
\end{abstract}


%

\input{chapters/1_introduction}
\input{chapters/2_related_work}
\input{chapters/3_algorithm}
\input{chapters/4_experiments_results}
\input{chapters/5_conclusion}

\bibliographystyle{splncs03}
\bibliography{Held}

\end{document}

%% file: chapters/0_abstract.tex
Community and cluster detection is a popular field of social network analysis.
Most algorithms focus on static graphs or series of snapshots.

In this paper we present an algorithm, which detects communities in dynamic graphs.
The method is based on shortest paths to high-connected nodes, so called hubs.
Due to local message passing we can update the clustering results with low computational power.

The presented algorithm is compared with other for some static social networks. The reached modularity is not as high as the Louvain method, but even higher then spectral clustering.
For large-scale real-world datasets with given ground truth, we could reconstruct most of the given community structure.
The advantage of the algorithm is the good performance in dynamic scenarios.

%% file: chapters/1_introduction.tex

\section{Introduction}
\label{sec:introduction}

Social network analysis has become very popular in the last years.
One part of this scientific field is cluster and community detection.
This could be used to describe changes in the network structure.
If we consider the famous Karate club example from \cite{Zachary1977}.
The community analysis from the member relation can describe, why the group has split up into two subgroups.

Most algorithms focus on a single analysis of a static graph, e.g.\ \cite{Donath1973, Macqueen1967}.
The next step is to use these algorithms on several snapshots of the same graph.
Changes in the clustering could be tracked with different algorithms, e.g.\ \cite{Takaffoli2011}.
There is also work done on dynamic graphs, e.g.\ \cite{Falkowski2007, Falkowski2008}.

In this paper we present an online capable algorithm to find communities based on high-connected hubs.

The paper is structured as follows. First, we give a brief introduction in \autoref{sec:related_work} into cluster and community structure and into related algorithms. Next, we present the proposed algorithm in \autoref{sec:algorithm} and experiments in \autoref{sec:experiments}. The paper will end with an conclusion in \autoref{sec:conclusion}.

%% file: chapters/2_related_work.tex

\section{Related Work}
\label{sec:related_work}

In this section, we introduce the term of cluster and community structure.
Also, we present some properties of social networks.
In the second part of this section we give a brief introduction into related algorithms.
These algorithms are used to compare the results of our algorithm.

\subsection{Cluster- and Community Structure}
\label{sub:cluster_community}

\begin{figure}
\centering
\includegraphics[width=\columnwidth]{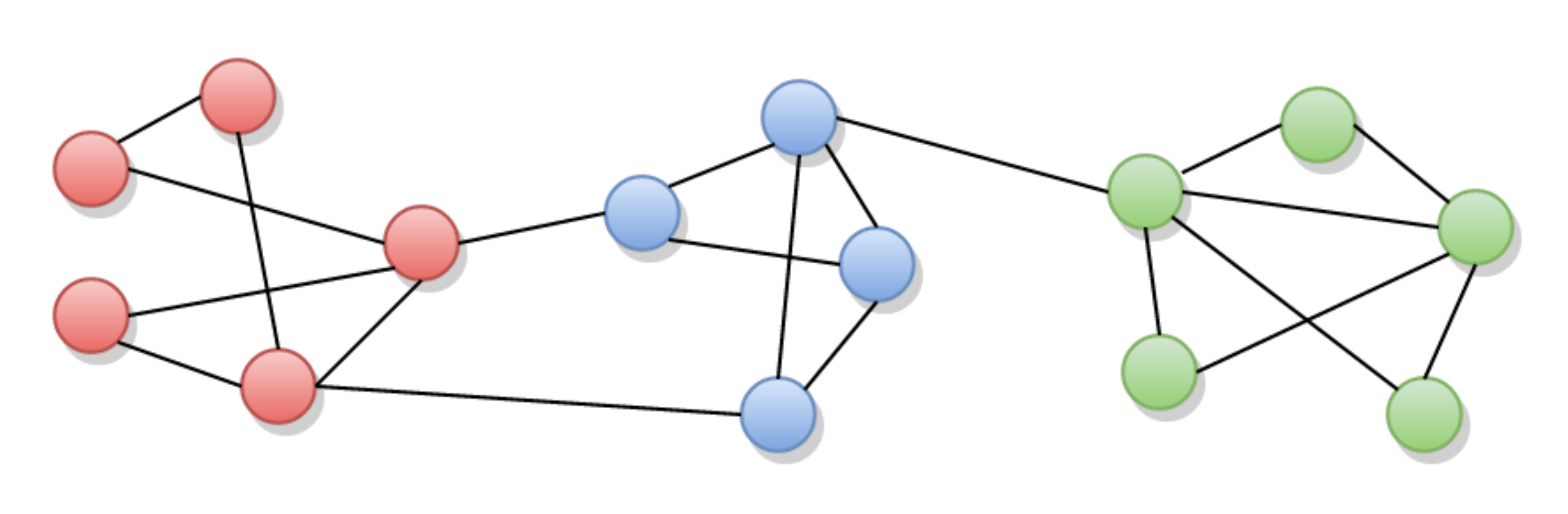}
\caption{Cluster structure}
\label{fig:cluster-structure}
\end{figure}

Social networks have the property, that they are clustered.
This means, that there are nodes, which are connected in a more dense way then other nodes, e.g.~\autoref{fig:cluster-structure}.

The main question is, how to find these clusters.
In social network analysis we can distinguish between a graph partition and covering.
The partition is related to the cluster structure. 
Every node is assigned to exactly one cluster.
In the more advanced covering communities (or fuzzy partitions) can overlap, so a node can be assigned to more then one community at a time.
This fact makes it hard to evaluate the quality of the community structure.

Two simple measures for evaluating the quality of a partition are the intra-cluster density and the inter-cluster sparseness. 
The intra-cluster density describes the ratio of existing and possible edges within the clusters.
This measure should be high, so clusters are connected strong.
The inter-cluster sparseness describes the ratio of existing and possible edges between nodes of different clusters. This value should be small, to get good separations between the clusters \cite{Fortunato2010}.

Another very popular measurement is the q-modularity proposed by Newman and Girvan \cite{Newman2004}.
The basic idea is, that random graphs have no cluster structure.
A comparison between the observed cluster density within a cluster should be higher then the density in these random graph.
If we maximize the q-modularity we come up with a good clustering.

In 1999, Barab\'asi et al.\ \cite{Barabasi1999} introduced the idea of scale free networks.
This means that node degree distribution of all nodes follows a power-law function.
\[
	P(k) \sim k^{-\gamma},
\]
where $P(k)$ is the probability of the degree $k$ of a node. Usually $2 \leq \gamma \leq3$.

From this distribution we get a lot of nodes with a low node degree and only less nodes with higher degrees.
These highly connected nodes are called hubs. \label{keyword:hubs}
In our algorithm, we use these hubs as starting points for the clusters.

\subsection{Related Algorithms}
\label{sub:related_algorithms}

\subsubsection{Spectral Clustering}

Spectral clustering \cite{Shi2000} uses an eigenvalue analysis of the normalized Laplacian matrix.
As input, we use the connectivity matrix of our graphs.
The first $k$ eigenvectors are used for a dimension reduction.

On the lower-space data, we use either k-means \cite{Ng2002} as partition algorithm or the discretization proposed by Yu and Shi~\cite{Yu2003}.

\subsubsection{Louvain Method}

In 2008 Blandel et al.~\cite{Blondel2008} propose an algorithm to extract community structure from large datasets.
The method is based on heuristics and modularity optimization.
The main idea is, that first each node is assigned an individual community.
Then iteratively, for every node a modularity gain for switching to adjacent communities is calculated.
If there is a positive modularity gain the node will put into the corresponding community with the highest modularity gain.
This process is done until no further increment is possible.

Now, there is a new graph build on the community structure of the input graph.
Each community becomes a node.
The edges in the new graph are representing the sum of edges between two communities.

The procedure will be repeated until there is a complete hierarchy of communities with only two communities on top.


%% file: chapters/3_algorithm.tex

\section{Algorithm}
\label{sec:algorithm}

Our algorithm to find clusters in dynamic graphs is based on two steps.
First, we have to determine all hubs in the network, which will be used as cluster centers.
Starting from them, we can propagate shortest paths to these hubs trough the whole network.
If the graph is altered, the clustering can be updated locally.

In the following we start with the selecting of hub nodes.
Afterwards, we present the basic algorithm.
Additionally, we propose some optimizations.

\subsection{Selecting Hubs}

As mentioned in \autoref{keyword:hubs}, hubs are highly connected nodes.
The simplest method is to determine the top $n$, relative or absolute, connected nodes and use them as hubs.
The main drawback of this method is, that all hubs has to be tracked and a decision if another node gets a hub could be done only with this tracked hubs.

To decide whether a node is a suitable hub individually, we introduce a threshold $d_{min}$ as minimal node degree.
Every node with $\deg(node) \ge d_{min}$ will be marked as an hub node.
This has the advantage that not all hubs have to be known for decision, so the analysis can be done without having the full graph. 

The threshold $d_{min}$ could ether set manually by the user or could be estimated from the graph.
If we have a scale-free network, the fraction of all nodes with an degree of $k$ follows the power-law distribution 
\[
	P(k) \sim k^{-\gamma}.
\]
The $\gamma$ must be estimated from the network, e.g. by using the maximum likelihood method.
If the user defines a fraction of hubs $h$ in the networks, $d_{min}$ could be estimated as follows:
\[
	d_{min} = \argmax_x D_n(x) \wedge D_n(x) \ge h
\]
with $n$ is number of nodes in the network, and
\[
	D_n(x) = \sum_{k=d_{min}}^{n} P(k).
\]

\subsection{Basic Algorithm}

The algorithm is based on passing hub information through the network.
Important changes in the network structure are propagated to all relevant nodes.

Each node stores a hub information table $T$ with the tuple entries $(h, p, \alpha)$, where $h$ represents the corresponding hub and $p$ the parent node, with the shortest path to the hub. 
$\alpha$ represents a weight of this information. 
Additionally we store the hub distance $d$.

\subsubsection{The Message} 
$M_{x \to y}(T', d')$ sent from a node to the neighbor nodes contains the basic hub information table 
\[
	T' = \left\lbrace\left(h, \frac{\alpha}{\sum_i \alpha_i}\right) : (h,p,\alpha) \in T\right\rbrace.
\]
The distance is set to 
\[
	d' = d + \omega(x, y),
\]
where $\omega(x,y)$ is the weight of the $(x,y)$-edge.

\subsubsection{Processing Messages}
$M_{x \to y}(T', d')$ in the target node.
We focus on three different cases. First, if $d'<d$ the new node distance in lower then the current distance. 
The hub information table is set to the table from the message, where $p$ is set to the sending node.
Also the distance is updated to the new distance.

The second case is, that the message has the same distance value then the current one $d'=d$. In this case we removed all tuples concerning the sending node and append the new information.
\[
	T_{new} = \left( \bigcup_{(h, p, \alpha) \in T \wedge p \neq x} (h, p, \alpha) \right) \cup \left\lbrace (h, x, \alpha) : (h, \alpha) \in T' \right\rbrace
\]

In both cases, the new hub information table will be propagated to all other neighbors.

If $d' \geq d + \omega(y,x)$, the sender has a worse connection then the receiver. 
Then the table is not updated, but the current table is propagated to the sender, so the sending node can update its distances.

Otherwise the message is dropped, with no further steps.

\subsubsection{Altering the graph}

Changes in the graph structure are handled as follows: If a new edge $(x,y)$ is added, node $s$ sends an information message to $y$. 
Due to message processing ether $y$ will update its hub information or sends its information to $x$. 
Additional the new structure is propagated through the network.

Whenever an edge is removed, we have to check both nodes, if the current edge was the connection to the parent node. 
Associated hub information tuple have to be removed. 
If this clears the table, the distance is set to $d=\infty$.
Changes have to be propagated to all neighbors.

If a node is removed, also all connected edges has to be removed and processed. 
Pure node creation does not influence the structure and does not have to be handled.

\subsubsection{Defining new Hubs}
If a node $n$ gets higher connected and becomes a hub, the distance value is set to $d=0$ and the hub information table is set to $T=(n, nil, 1)$.
The information has to be propagated to all neighbors.

\subsubsection{Removing Hubs}
If a node $n$ loses his hub state, the distance is set to $d=\infty$ and the information table is cleared.
After propagating this information all neighbor nodes send alternative hub information.

\subsubsection{Assign Cluster or Community}
Based on the $\alpha$ value in the hub information table of each node, we can assign cluster and community labels.
If we want to have crisp cluster assignments, we sum up all $\alpha$ values for each $h$.
The $h$ with the highest sum, will be the cluster label.

If we do not need crisp assignments, we normalize the $\alpha$ sums and use them as membership degree for a certain community.

\subsection{Optimization}

\subsubsection{Running Initial Steps parallel}
We propose to collected all open messages in a priority queue, ordered by distance.
This yields into a breath first search around the hubs.
Especially if there are multiple new detected hubs at the same time, e.g.\ in the static scenario, where the whole network is given.
This can decrease the number of messages to be processed dramatically.

\subsubsection{Hash message -- avoid duplicates}
Another optimization step is the hashing of open message edges. 
Nodes that are on the border of two clusters get multiple information from both hubs.
These combined information have to be send to all following nodes.
So, if there is a message $M_{x \to y}$ in the queue and $x$ gets new information another message $M_{x\to y}$ is created and the old message could be removed.
This could be done by storing the message content on the edge, while having the not processed edges in the priority queue.

%% file: chapters/4_experiments_results.tex

\section{Experiments}
\label{sec:experiments}

We will do a two step evaluation of our presented algorithm, first a static and second a dynamic test.
In the static test, we will compare the clustering results with other algorithms.
The dynamic test should check the performance for dynamic graphs.

\subsection{Static Test}
\label{sub:test_settings}

First we will check the algorithm in a static test setting.
We take some well-known datasets to evaluate and compare clustering results with the spectral clustering and louvain method.

We will start with a deeper view on the well-known karate dataset from Zachary~\cite{Zachary1977}.
He observed the relation of the members of the club, after the club has split up into two groups.

\begin{figure}
\centering
\includegraphics[trim=3cm 2cm 3cm 2cm, clip=true, width=0.8\columnwidth]{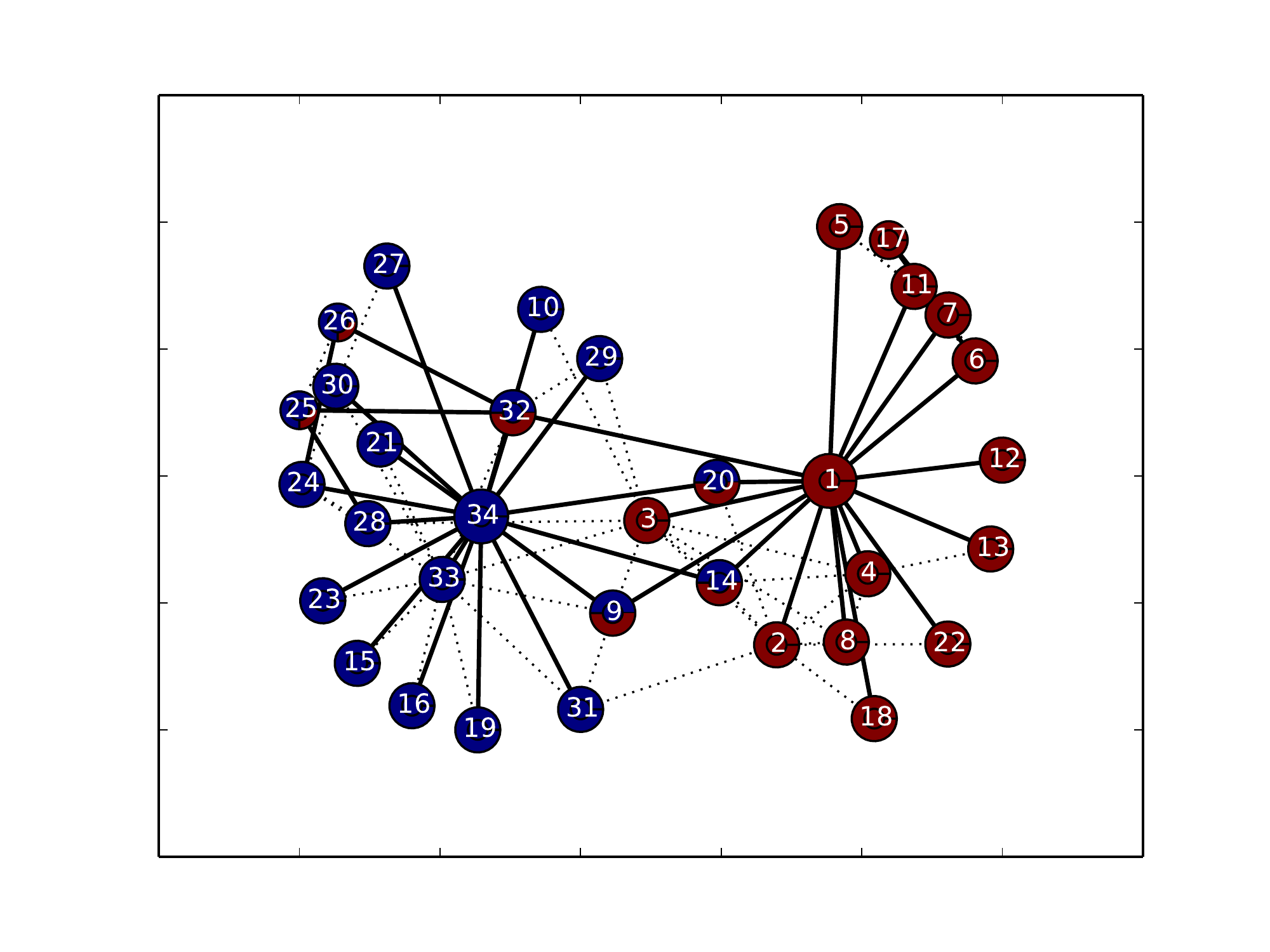}
\caption{NHC-clustering: karate club}
\label{fig:nhc-karate}
\end{figure}

\autoref{fig:nhc-karate} shows the results of the clustering with our algorithm.
The outer circle describes the community membership distribution of the different communities.
The inner circle is the crisp cluster association described in \autoref{sec:algorithm}.
The solid lines describe connections, which are used for next hub propagation.
The dotted lines are irrelevant for the algorithm and could be removed without need to do further processing steps.

In the figure we can see the two cluster center nodes 1 and 34.
The nodes 9, 14, 20, and 32 are exactly in-between the two center nodes.
Due to crisp partition they are associated to cluster 34, but they could also be assigned to cluster 1.

\begin{figure}
\centering
\includegraphics[trim=3cm 2cm 3cm 2cm, clip=true, width=0.8\columnwidth]{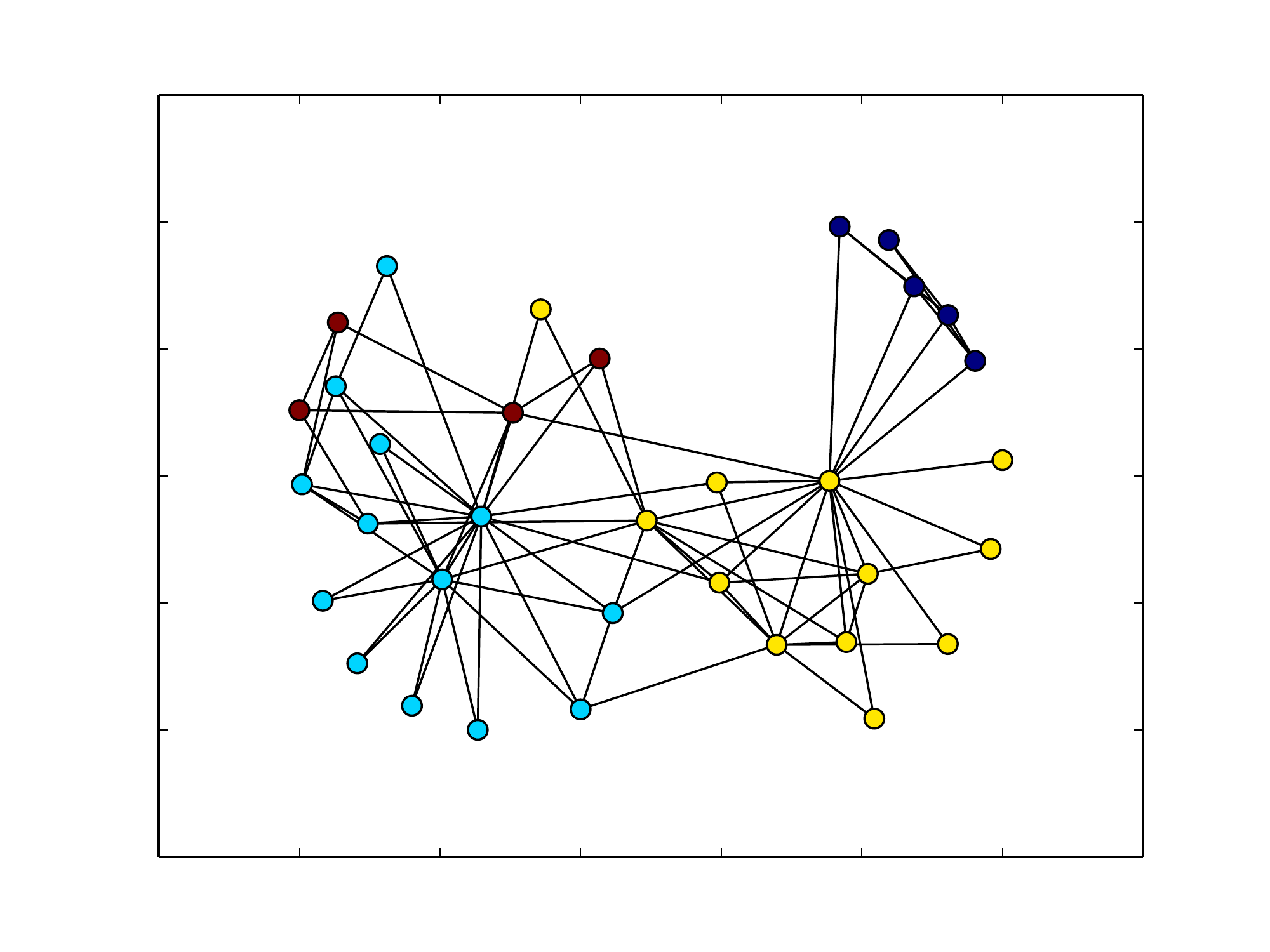}
\caption{Louvain-clustering: karate club}
\label{fig:louvain-karate}
\end{figure}

\autoref{fig:louvain-karate} shows the same graph clustered with the louvain method.
This method offers four clusters, where the two main clusters are similar to the NHC results.

\begin{figure}
\centering
\includegraphics[width=0.8\columnwidth]{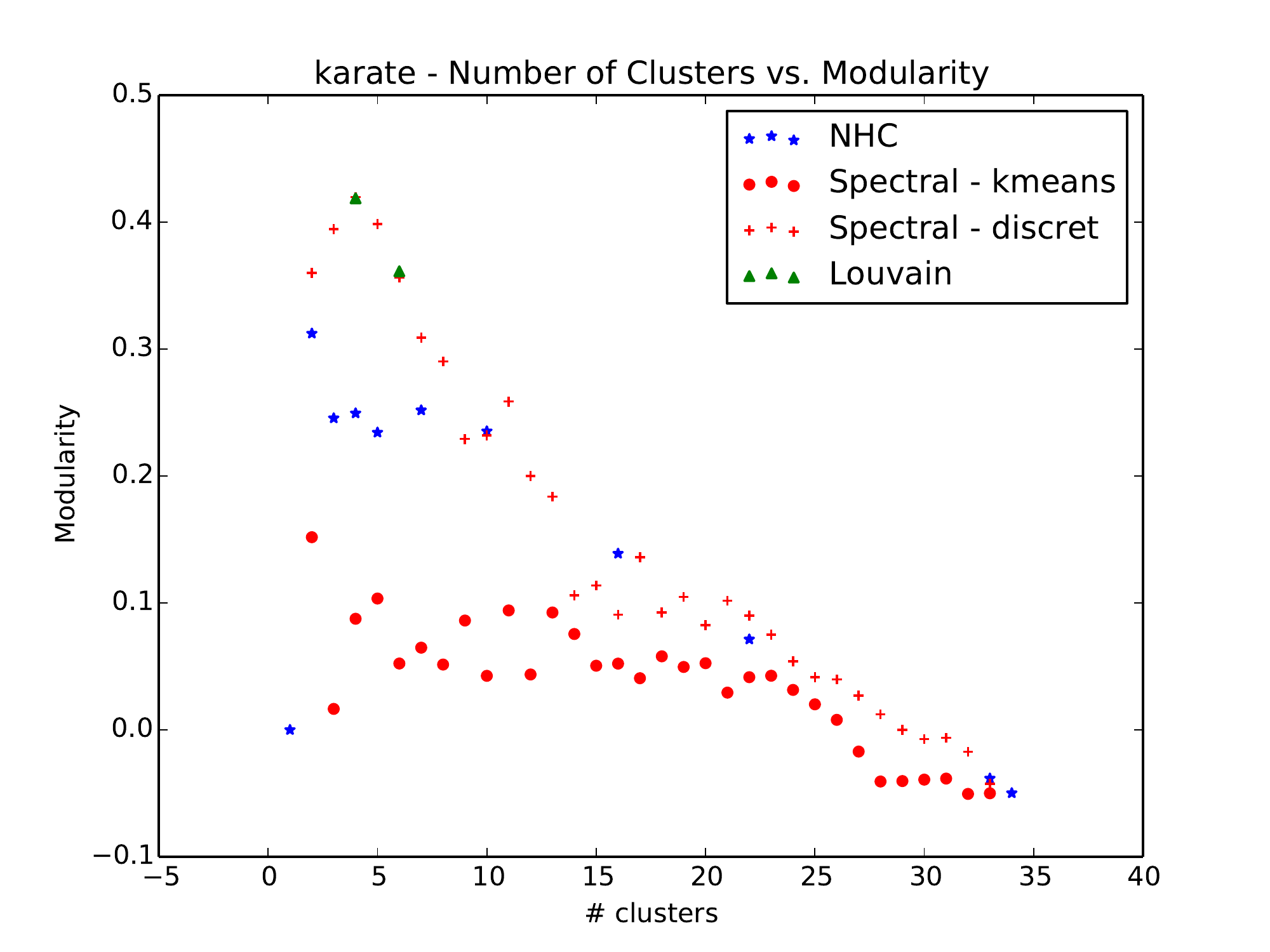}
\caption{Modularity comparison for different number of clusters (karate dataset)}
\label{fig:mod-karate}
\end{figure}

In \autoref{fig:mod-karate} we show the reached modularity for NHC, spectral clustering, and the louvain method for different number of clusters.
The louvain method only offers results for four and six clusters. 
A more detailed structure is not possible.
NHC outperforms the spectral clustering with k-means for every number of clusters.
Especially for small number of clusters the spectral clustering with discretization and the louvain method produce good results for the modularity.
For 10 and more clusters NHC produces similar results to discretized spectral clustering.
The lower values for lower number of clusters could be caused by the assignment method for equal distributed memberships.

\begin{figure}
\centering
\includegraphics[width=0.8\columnwidth]{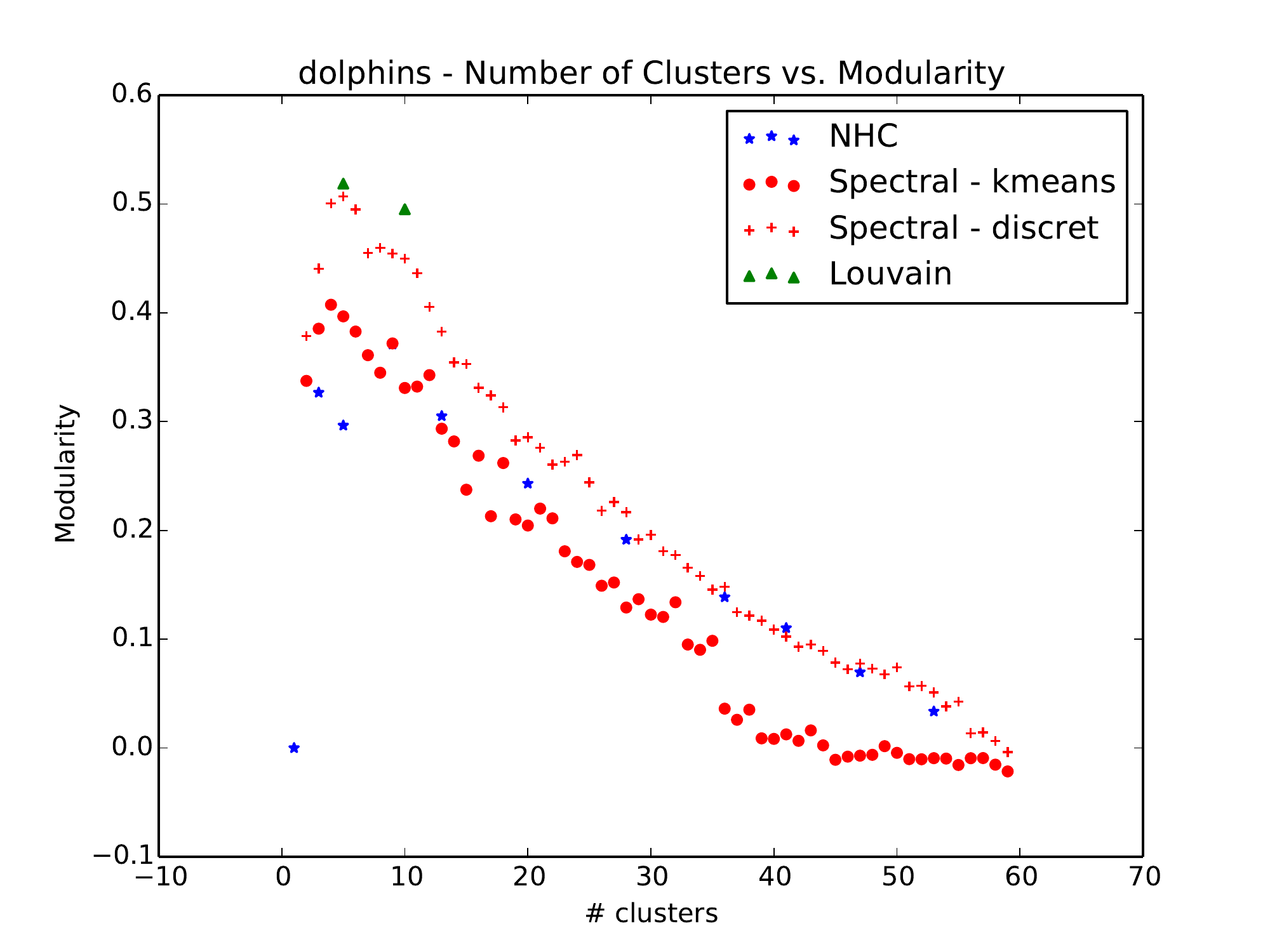}
\caption{Modularity comparison for different number of clusters (dolphins dataset)}
\label{fig:mod-dolphins}
\end{figure}

In \autoref{fig:mod-dolphins} we can see the modularity for the dolphins network~\cite{Lusseau2003}. 
The dataset represents the social structure of 62 dolphins.
NHC does not reach the modularity of the other methods for lower number of clusters, but for higher numbers the values are similar to spectral clustering with discretization.
Again the louvain method could not produce results for finer structures.

\begin{figure}
\centering
\includegraphics[width=0.8\columnwidth]{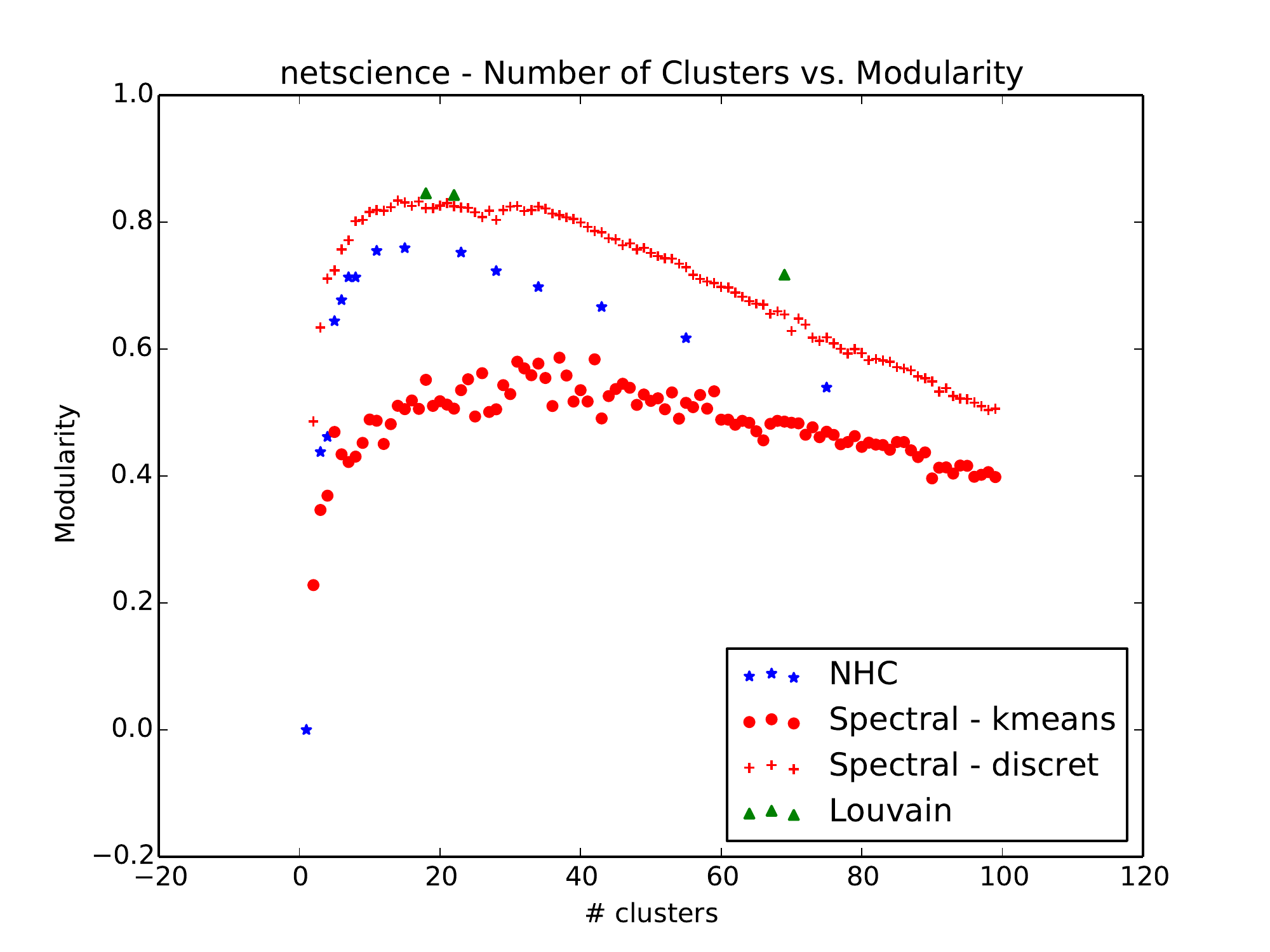}
\caption{Modularity comparison for different number of clusters (netscience dataset)}
\label{fig:mod-netscience}
\end{figure}

\begin{figure}
\centering
\includegraphics[trim=10cm 7cm 10cm 6.5cm, clip=true, width=.8\textwidth]{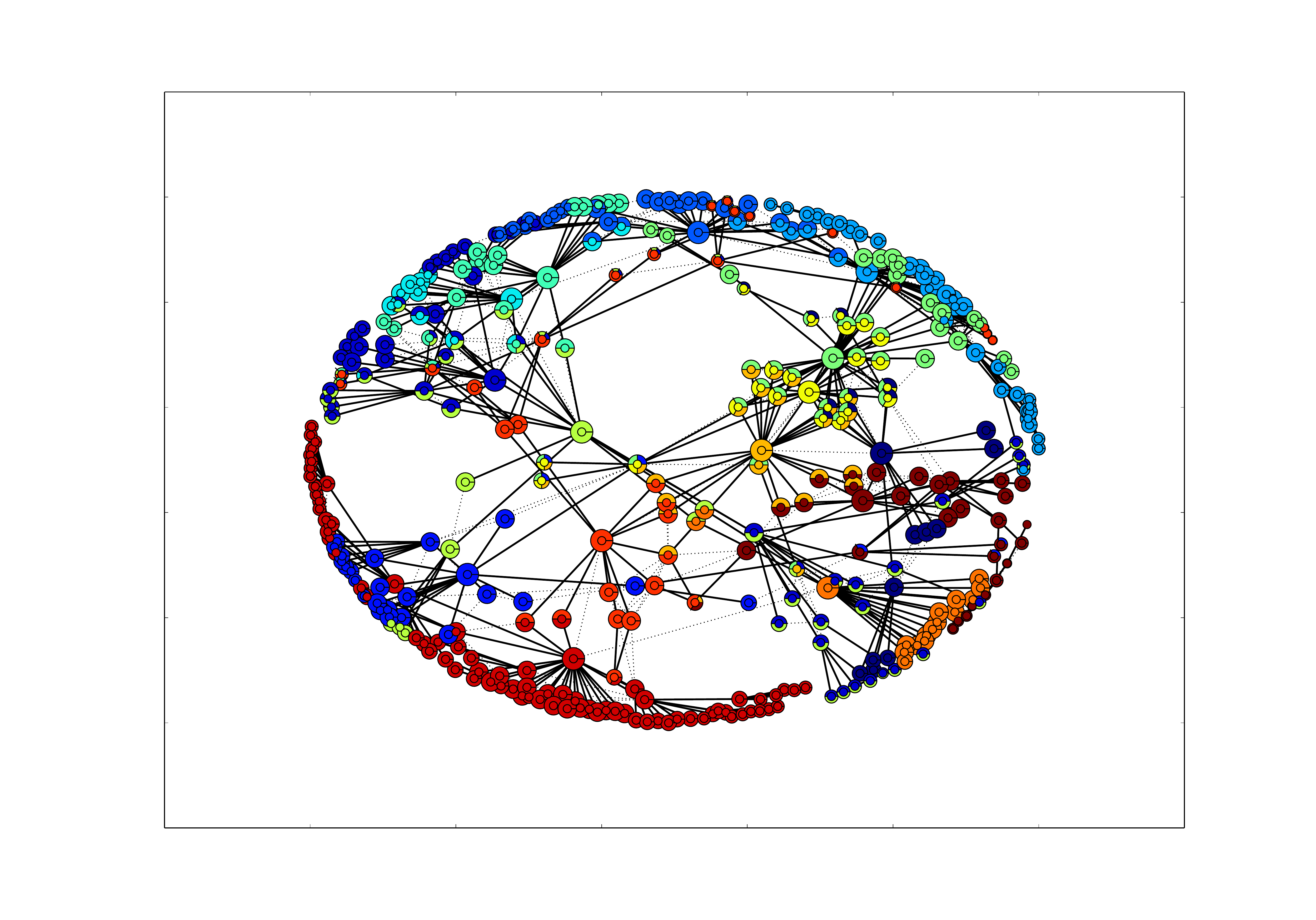}
\caption{NHC-clustering: netscience}
\label{fig:nhc-netscience}
\end{figure}

\begin{figure}
\centering
\includegraphics[trim=10cm 7cm 10cm 6.5cm, clip=true, width=.8\textwidth]{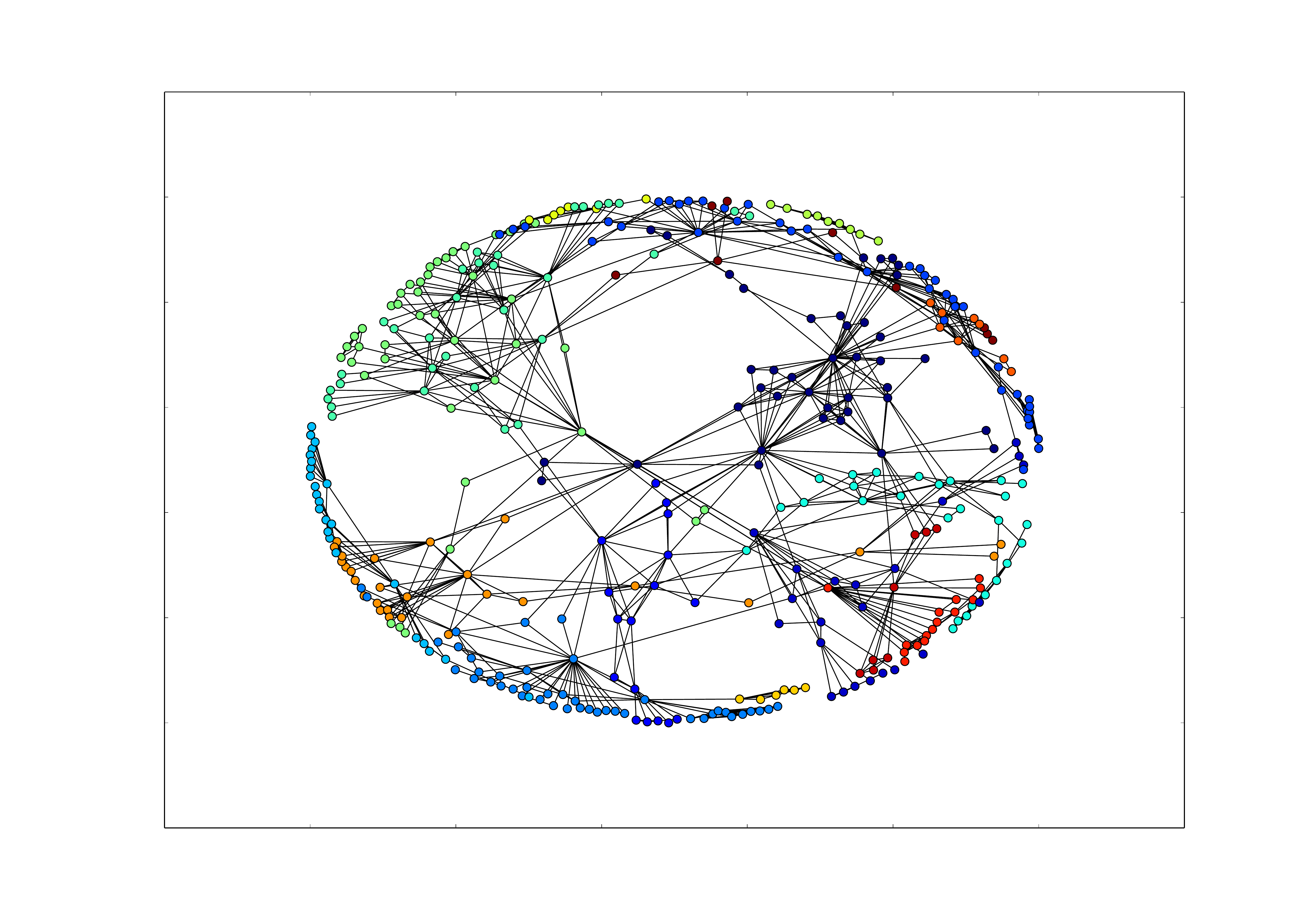}
\caption{Louvain-clustering: netscience}
\label{fig:louvain-netscience}
\end{figure}

Our last dataset is the netscience dataset assembled by Newman~\cite{Newman2006}.
The dataset contains a coauthorship network of scientists.
The network contains 1589 nodes, with weighted edges.
We focus on an unweighted version.
In \autoref{fig:nhc-netscience} you will find the cluster results from our algorithm.
A comparison to the louvain-clustering in \autoref{fig:louvain-netscience} shows a high similarity in cluster assignments of the nodes.
The reached modularity by our algorithm is lower then the louvain method, and also lower then spectral clustering with discretization, but even higher then spectral clustering with k-means.

\newpage

\subsubsection{Large Scale Networks}

Yang and Leskovec~\cite{Yang2013} provide a set of large scale online communities including a ground truth.
We used there the DBLP, YouTube, and Amazon dataset to check the cluster prediction performance of our algorithm. These datasets contain from $300,000$ to $1,100,000$ nodes and from $925.000$ to almost $3,000,000$ edges. In the dataset ground truth, they found from $8,300$ to $75,000$ communities.

We use the V-measure-score~\cite{Rosenberg2007} to evaluate the cluster performance.
Due to the computational complexity of this measure, especially for large amount of clusters, we will use the top 5000 communities provided by Yang and Leskovec.
The selection was done by different community measures.

Due to the complexity of spectral clustering and the fact we have to check a lot of different number of communities we skip this algorithm for this experiment. We will compare Louvain results with our proposed algorithm.

\begin{table}[htb]
\centering
\caption{Large scale network comparison for Louvain and Nearest Hub Clustering}
\label{tab:large}
\begin{tabular}{p{5cm}ccc}
\toprule
~ & DBLP & Amazon & YouTube \\
\midrule
Number of nodes & $ 317,080 $ & $ 334,863 $ & $1,134,890$ \\
Number of edges & ~ $1,049,866$ ~ & ~ $925,872$ ~ & ~ $2,987,624$ ~ \\
Number of communities & $13,477$ & $75,149$ & $8,385$ \\
Top 5000 communities - nodes & $ 112,228 $ & $ 67,462 $ & $72,959$ \\
\midrule
Louvain - time & \unit[98,7]{s}&  \unit[81.7]{s} & \unit[252,2]{s}\\
Louvain - V-measure-score & $0.525$ & $0.863$ & $0.450$\\
Louvain - NMI-score & $0.530$ & $0.871$ & $0.510$\\
Louvain - q-modularity  & $0.818$ & $0.926$ & $0.710$\\
\midrule
NHC - time & \unit[135.35]{s}& \unit[168.31]{s} & \unit[154,8]{s}\\
NHC - min-degree &12& 16& 4\\
NHC - V-measure-score & $0.726$ & $0.944$ & $0.832$\\
NHC - NMI-score & $0.746$ & $0.945$ & $0.842$\\
NHC - q-modularity  & $0.432$ & $0.613$ & $0.297$\\
\bottomrule
\end{tabular}
\end{table}

From Table~\ref{tab:large} we get, that the Louvain method gets much higher values for q-modularity, but this is not correlated to better community structure given by the ground truth from the datasets.
We got V-measure-scores from $0.73$ to $0.944$ which indicates that a large amount of community structure could be found by our algorithm.

\subsection{Dynamic Test}
\label{sub:dynamic_test}

In this section we will check the dynamic behavior of the presented algorithm.
We generate random clustered powerlaw-graphs with the algorithm proposed by Holme and Kim~\cite{Holme2002}.
As parameters we choose $n=1000$ nodes, which are connected each to $m=10$ other nodes.
With a probability of $p=0.7$ the model will create a triangle to increase the cluster coefficient of the resulting network.

\subsubsection{Adding Edges}

\begin{figure}
\centering
\includegraphics[width=0.8\columnwidth]{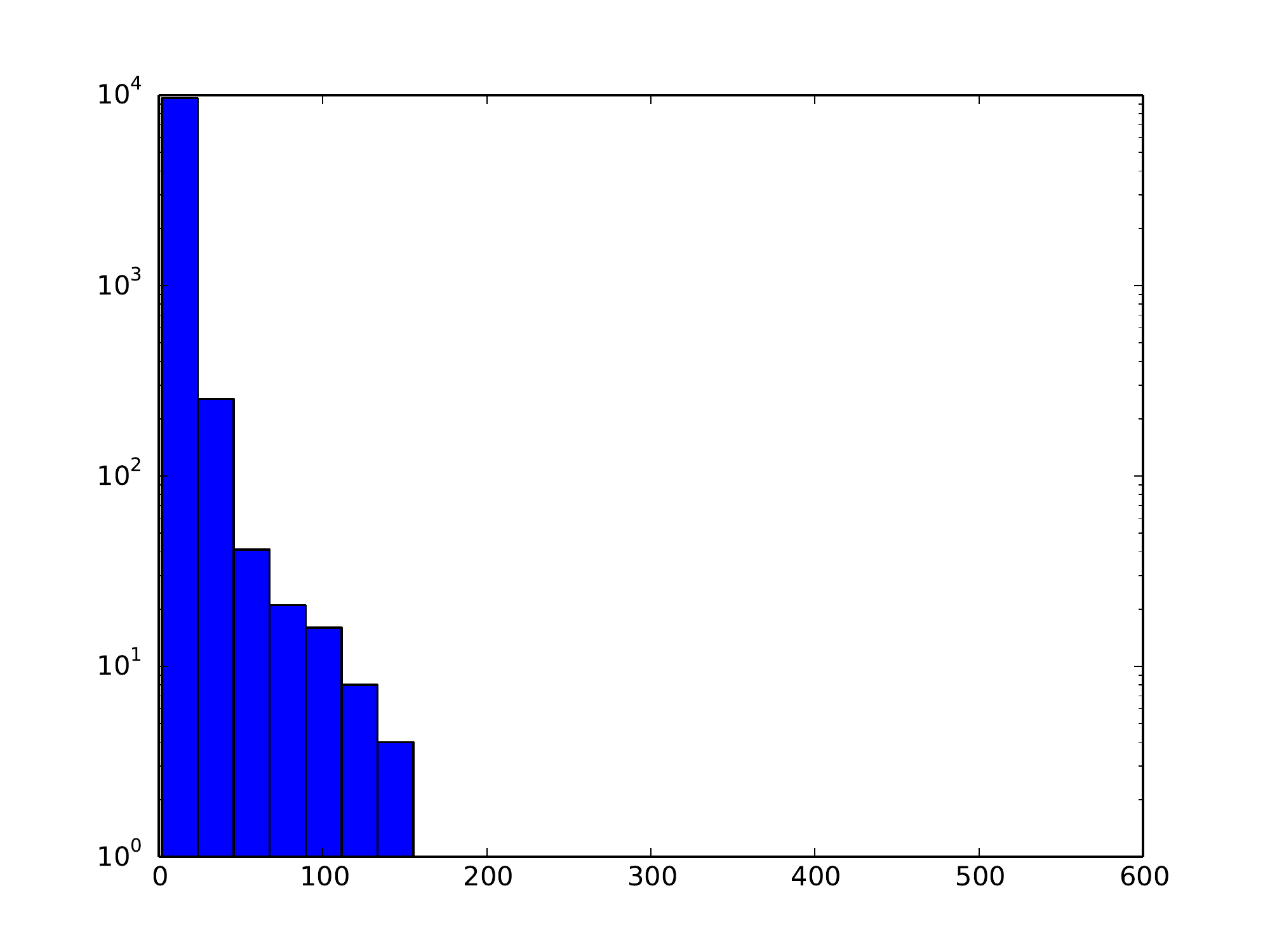}
\caption{Logarithmic histogram of messages send during 10000 adding events.}
\label{fig:add-hist}
\end{figure}

To test the adding edges behavior we generated 100 graphs and added 100 random edges to each graph.
This yields into 10000 adding edged events.
In \autoref{fig:add-hist} we show a logarithmic histogram of the processed messages distribution.
If a new edge does not create a new shorter path to the hubs, which happens in \unit[63.3]{\%} of the cases, only two messages between the two new connected nodes have to be processed.
On average 7.73 messages and maximal 549 messages have been send.

Due to the low amount of messages needed to update the clustering during adding edges to the network, the presented algorithm performs well in dynamic adding nodes and edges.

\subsubsection{Removing Edges}

\begin{figure}
\centering
\includegraphics[width=0.8\columnwidth]{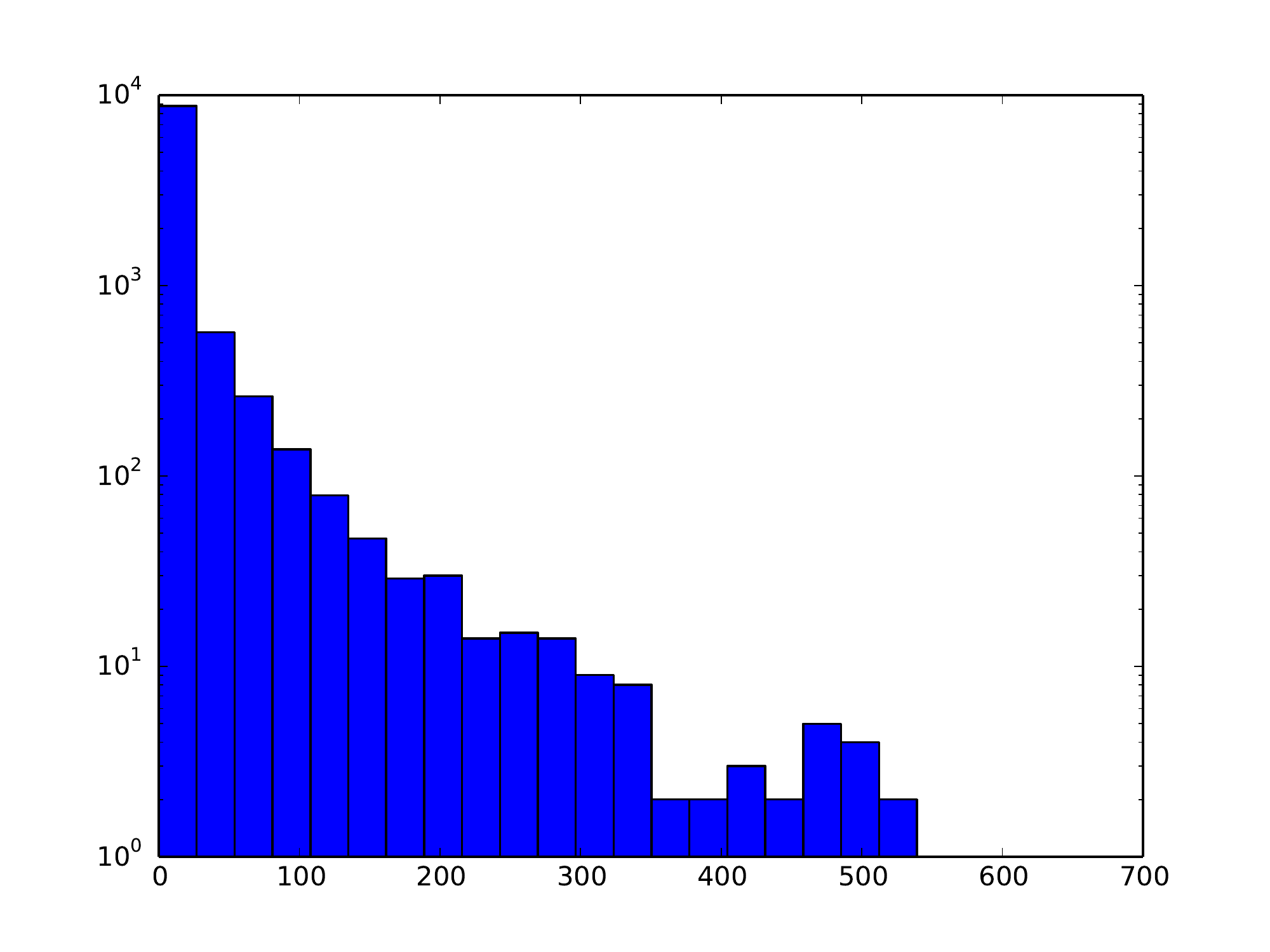}
\caption{Logarithmic histogram of messages send during 10000 removing events.}
\label{fig:remove-hist}
\end{figure}

We performed similar test for removing edges.
Again we created 100 graphs and removed randomly 100 existing edges from each graph, so we get 10000 removing events.
In \autoref{fig:remove-hist} is a logarithmic histogram of the send messages distribution presented.
If the removed edge does not destroy the shortest path to an hub, we do not have to send any messages at all, which happens in \unit[60.1]{\%} of the events.
On average we need 14.2 messages. Maximal 674 messages have send, this is a really rare case where the structure near the hubs is changed.

Also the removing process works with a low amount of send messages.
This shows, that the presented algorithm performs well on dynamic graph structures.

%% file: chapters/5_conclusion.tex

\section{Conclusion}
\label{sec:conclusion}

We presented an algorithm for graph clustering.
In static tests it shows that the algorithm produces high modularity results for finer structure.
In contrast to the louvain method or the spectral clustering with discretization the algorithm did not reach the global optima of modularity.

Tests on large-scale real-world datasets show, that large parts of the underlying community structure could be found by our algorithm.
A V-measure-score analysis again a ground truth given by~\cite{Yang2013} showed scores from $0.73$ to $0.944$. This is much more then the Louvain method reached.

The main advantage of the algorithm is the dynamic behavior.
If the graph changes over time, only a small amount of processing steps have to be done to update the clustering.
This enables the it for online cluster and community analysis.

The algorithm itself generates overlapping communities and provides membership degrees.
These results are deterministic.
Randomness influences only the crisp cluster assignment.

The next steps we focus on is the optimization of the hub threshold.
An implementation in python will be provided at \url{http://bitbucket.org/paheld/dynamix}.